# The precursory electric signals, observed before the Izmit Turkey EQ (Mw = 7.6, August 17th, 1999), analyzed in terms of a hypothetically pre-activated, in the focal area, large scale piezoelectric mechanism.

Thanassoulas[1], C., Klentos[2], V.

1. Retired from the Institute for Geology and Mineral Exploration (IGME), Geophysical Department, Athens, Greece.
   e-mail: thandin@otenet.gr - URL: www.earthquakeprediction.gr

2. Athens Water Supply & Sewerage Company (EYDAP),
   e-mail: klenvas@mycosmos.gr - URL: www.earthquakeprediction.gr

**Abstract**

The generated, prior to the Izmit Turkey large EQ, preseismic electric signals were recorded in Greece by the **VOL** Earth's electric field monitoring site. In order to explain their peculiar character and their generating mechanism, a large scale piezoelectric mechanism was assumed that was initiated in the Izmit seismogenic region long before the EQ occurrence time. The theoretical analysis of the adopted physical model justifies the generation of a number of specific electric signals that can be emitted from the focal area before the rock formation failure. The processing of the registered by the **VOL** monitoring site raw data revealed the presence of similar signals as the expected theoretical ones. Therefore, it is concluded that long before the Izmit EQ occurrence a large scale piezoelectric mechanism was initiated that was modulated too by the tidally triggered lithospheric oscillation and therefore generated the observed preseismic electric signals. The adopted piezoelectric model provides critical information about the time of occurrence of the seismogenic area rock formation failure and therefore the possibility for a real short-term time prediction of a large EQ. The other two predictive EQ parameters, location and magnitude, are discussed in the frame of electric field triangulation and the Lithospheric Seismic Energy Flow Model (LSEFM).

## 1. Introduction.

On Tuesday, August 17th, 1999, at 3:02 a.m. local time, an earthquake of Mw = 7.6 struck the Kocaeli province (near Izmit town) of north-western Turkey. The earthquake occurred at the western branch of the 1,500-km-long North Anatolian Fault (NAF) system which was ruptured at a length of about 110 km. According to official Turkish government estimates, the earthquake caused 17,127 deaths, 43,953 injuries, more than 250,000 people were displaced and a large number of damages in buildings and large scale technical infrastructure.

Details for the geological and seismotectonic environment of the regional Western part of the North Anatolian Fault zone, where the EQ occurred, have been presented by many researchers (i.e. Barka 1999; Ferrari et al. 2000; Nalbant et al. 1998; Stein et al. 1997; Taymaz 1999; USGS 1999). The vast majority of the research literature on this topic is directly (or in a way indirectly) seismically oriented.

In this work, the Izmit EQ will be studied in a quite different point of view. Its time of occurrence will be correlated to the observed preseismic electric signals, that were generated long before its occurrence time, due to excess strain load at its focal area and were recorded by the **VOL** monitoring site located in Greece at a distance of 650 Km from the EQ epicentral area. The location of **VOL** monitoring site and the Izmit EQ are presented in figure (1).

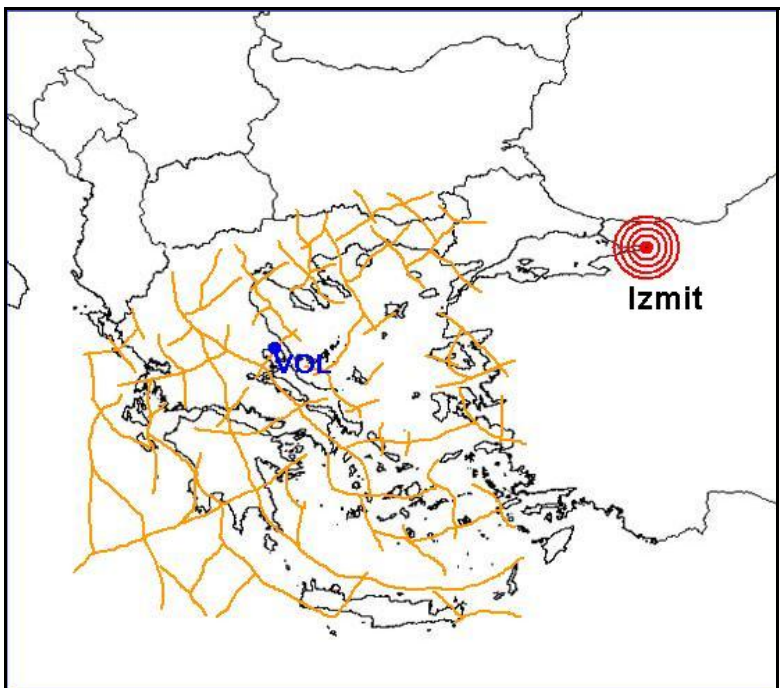

Fig. 1. Location of **VOL** monitoring site (blue solid circle) and the IZMIT EQ (red circles). The thick brown lines represent the deep lithospheric fracture zones and faults of the Greek territory (Thanassoulas, 2007).

Moreover, a physical mechanism will be postulated that accounts for the generation of the observed preseismic electric signals and the time of occurrence (in short-term mode prediction) of the Izmit EQ.

**2. Theoretical analysis - postulated model.**

The Earth-tides mechanism was early recognized as a potential trigger for the occurrence of strong earthquakes. This approach was followed to study the time of occurrence of EQs and their correlation to Earth-tides. Knopoff (1964), Shlien (1972), Heaton (1982) and Shirley (1988) suggested the Earth-tides as a triggering mechanism of strong EQs, Yamazaki (1965, 1967), Rikitake et al. (1967) studied the oscillatory behaviour of strained rocks, due to Earth-tides, while Ryabl et al. (1968), Mohler (1980) and Sounau et al. (1982) correlated Earth-tides to local micro-earthquakes and aftershock sequences.

The following figure (2) demonstrates the mechanism which generates the Earth-tides. On the left, the forces are shown which apply the Moon or the Sun on the Earth's surface and the corresponding generated lithospheric deformation. A typical daily tidal variation is shown on the right.

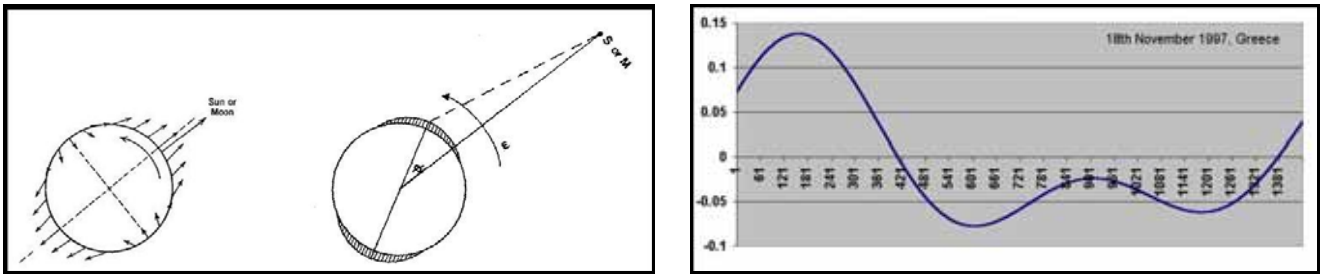

Fig. 2. Left: Forces applied on the Earth's surface (Stacey, 1969) by the Moon or the Sun (left) and the resulted, deformation (Garland, 1971) of the lithospheric plate. Right: Earth-tide calculated for the 18$^{th}$ November, 1997, Greece $\varphi = 38^0$, $\lambda = 24^0$. The vertical scale is in mgals, while the horizontal one is in minutes (1 day = 1440 minutes). In this figure it is possible to identify the **K1** (23.93hr), **K2** (11.97hr) and **S2** (12.00hr) components.

Earth-tides exhibit an oscillating mode of behaviour. The study of the Earth-tide oscillation has shown that, basically, it consists of the following main components, presented in the table below.

**Tidal components**

| Symbol | Name | Period (hr) |
| --- | --- | --- |
| M2 | Principal Lunar | 12.42 |
| S2 | Principal Solar | 12.00 |
| N2 | Lunar Ellipticity | 12.66 |
| K2 | Lunisolar | 11.97 |
| K1 | Lunisolar | 23.93 |
| O1 | Lunar Declination | 25.82 |
| P1 | Solar Declination | 24.07 |
| M1 | Moon declination | 14 days |
| $S_{sa}$ | Moon declination | 6 months |

Finally, an Earth-tide wave is generated with a year's period, because of the Earth's motion in an ellipse with the Sun in the focus.

The deformation of the lithosphere follows the oscillatory character of the Earth-tides. Garland (1971), Stacey (1969), Sazhina and Grushinsky (1971), study in detail this type of Earth's oscillation, not to mention the majority of the Geophysical textbooks. A detailed presentation of the generation of the normal to the ground oscillatory tidal component was given by Thanassoulas (2007) while a brief presentation (key-drawings) is given below in figure (3).

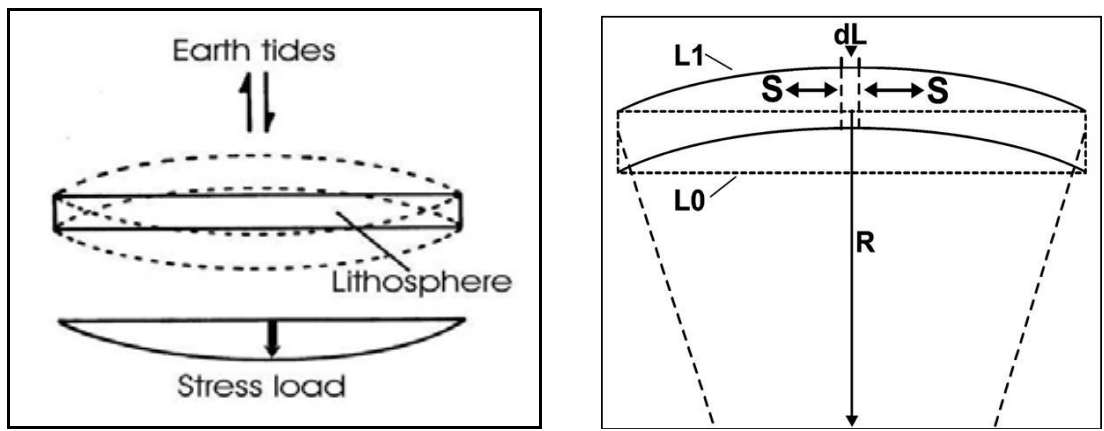

Fig. 3. Left: Lithospheric plate at rest (solid line) and maximum amplitude oscillation (dashed line). Bottom drawing indicates the stress-load, along the plate, during its oscillation. Right: Compressional and extensional oscillatory forces (**S**) are generated in the lithosphere, due to its tidal oscillation (for more details see: Thanassoulas, 2007).

The oscillation of the lithospheric plate, due to Earth-tides, has two severe consequences:

-**The first one** is that, at the peak amplitude of its oscillation, when the plate is at nearly critical stress load, <u>it can reach the extreme conditions which are necessary for an earthquake to occur</u> (Thanassoulas et al. 2001), provided that, the focal area itself, has been charged, enough, by the linear increase of stress, due to plate's motion. This takes place, especially, when all the oscillating components of the Earth-tides are "in phase".

-**The second** one is that, at the very same stress load critical conditions, the focal area of an imminent earthquake generates electrical signals (Thanassoulas 1991, Clint 1999). The latter is due to its large crystal lattice deformation and due to the piezoelectric properties of the quartzite content of the lithosphere. As a consequence, a large scale piezoelectric mechanism in the regional seismogenic area can be activated.

The type of electric signals, which are expected to be observed after the activation of the piezoelectric mechanism, cover a wide frequency spectrum due to the nature of the progressive lithospheric mechanical failure before and during the EQ occurrence. Some of these electric signals are of specific earthquake predictive interest.

The most well-known electric signal that is generated by piezoelectricity is the one presented in the following figure (4). The generated by the strained volume electric potential is almost linearly related to the volume strain as long as the applied strain level exceeds the piezoelectric mechanism activation strain level.

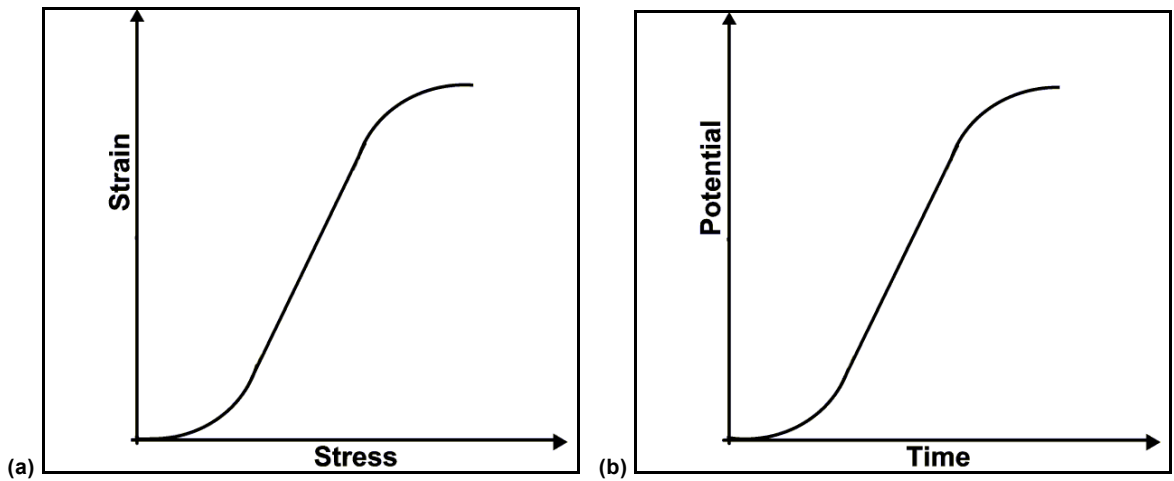

Fig. 4. Typical stress – strain (a) relation of solid material that exhibits piezoelectric-like properties and its corresponding (b) potential – time relation. It is assumed that positive charges are generated in compression mode.

Although the latter type of electric signal can be quite easily be observed in a laboratory experiment on a rock sample, it is rather impossible to directly observe it in situ on a seismogenic region. The problem is that a seismogenic area cannot be treated as a rock sample under laboratory conditions due to its large (in Km) dimensions. This problem can be resolved (Thanassoulas 2007, 2008) indirectly by integrating along time the registered gradient electric potential data obtained at large distances from the piezoelectrically (seismogenic) activated area.

The basic piezoelectric mechanism presented in figure (4) gives rise, indirectly, to the following types of signals. Let us use a low-pass potential gradient detecting system. In such a case it is possible to detect the first derivative of the piezoelectric potential generated over the seismogenic area (fig. 5 left). This is known as the Very Long Period (**VLP**) electric signal. If instead of a low-pass filter we use a band-pass filter then an oscillating potential (fig. 5 right) will be detected with a period similar to the oscillating stress component. A characteristic of this signal is that its amplitude increases as long as the strain increases. The latter is a direct outcome of the observed increase of the potential gradient as a function of stress.

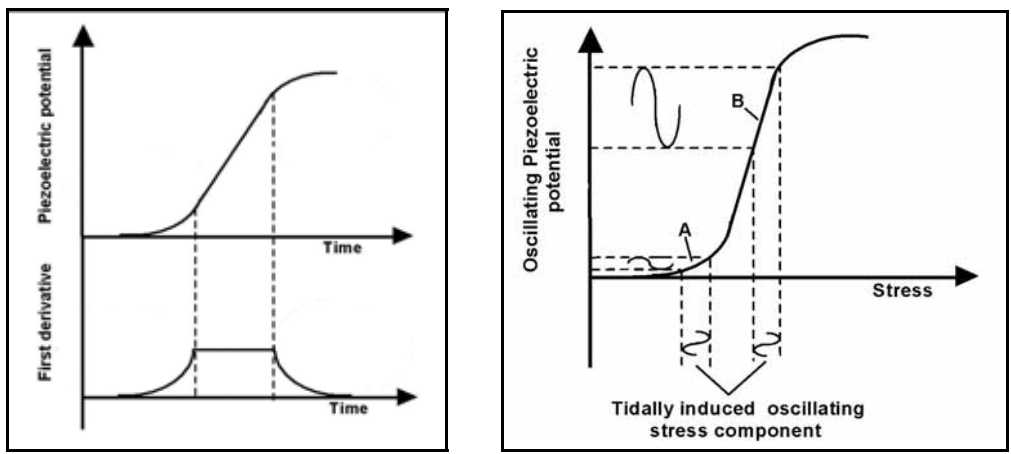

Fig. 5. Left: First derivative: **VLP** signals observed through a low-pass filter. Right: Oscillating signal generated due to oscillating stress component, observed through a band-pass filter (Thanassoulas et al. 1986, 1993).

The form of the potential – strain relation can generate short wave-length, compared to the duration of the piezoelectric activity, electric signals. In figure (6) left, in red circles are presented the two (**A, B**) non-linear parts of the potential – strain curve. These non-linear parts generate higher harmonic (compared to the basic one) electric signals. Furthermore, the piezoelectricity can be activated in a varying scale of rock volumes and therefore, short or long wave-length electric pulses can be generated by combining compresional and decompresional piezostimulated currents (Thanassoulas, 2008a). The latter mechanism is shown in figure (6) right.

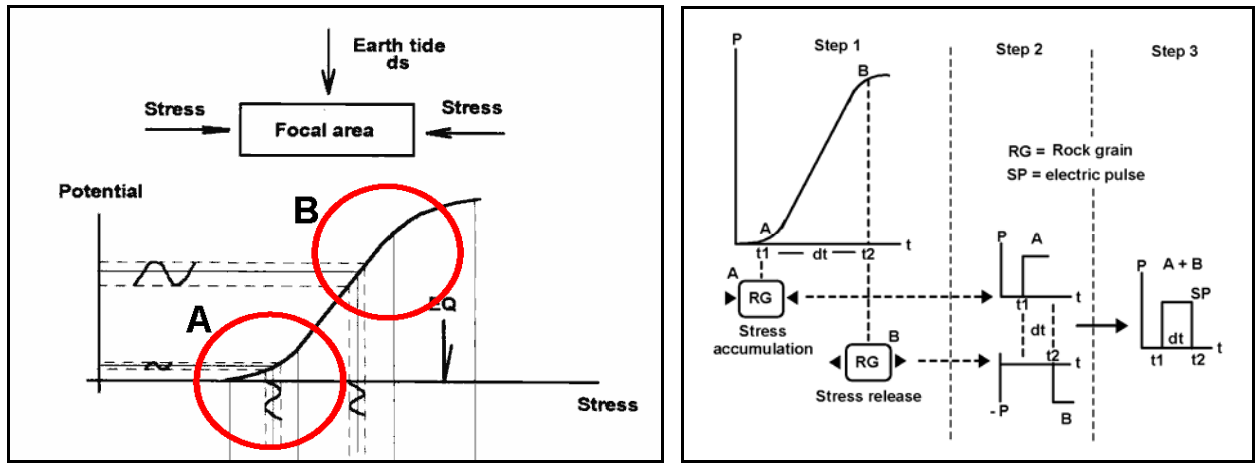

**Fig. 6.** Left: Non-linear areas (**A, B**) of the piezoelectric potential curve are shown, where electric signals of higher harmonics can be generated. Right: Schematic presentation of the generation of a square electric pulse by the mechanisms of piezoelectricity and piezostimulated currents generation. A rock small block / grain (**RG**) is subjected in (**A**) stress increase and stress decrease in (**B**) at step (1). Positive (**A**) or negative (**B**) currents generate at step (2) which are combined in step (3) in a short square pulse (**SP**) of $dt$ duration.

The already presented theoretical model will be applied on and tested against the electric potential registered at **VOL** monitoring site in Greece before the Izmit EQ occurrence time.

### 3. Data presentation and processing.

The electrical potential as a function of time, which was recorded at Volos (**VOL**) monitoring site, Greece (Thanassoulas et al. 2000) and includes the Izmit, EQ (17$^{th}$, August, 1999, M=7.6) occurrence time, is shown in the following figure (7). Figure (7) spans from June 20$^{th}$ to August 30$^{th}$, 1999. The recorded potential is the gradient in time of the generated by the piezoelectric mechanism total electric field (Thanassoulas 2007). The red horizontal bar represents the most "noisy" part of the recording while the vertical red bar indicates the occurrence time of the Izmit EQ.

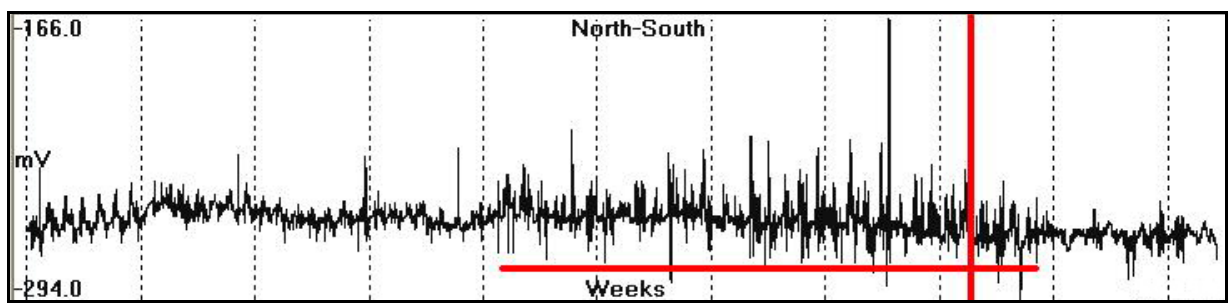

**Fig. 7.** Gradient in time of the piezoelectric potential (June 20$^{th}$ to August 30$^{th}$, 1999) generated by the activation of the piezoelectric mechanism in the seismogenic region of the Izmit EQ. The most "noisy" part of it is indicated by a horizontal red bar while the Izmit EQ occurrence time is indicated by a vertical red bar.

At a first glance it is evident that outstanding electric noise was generated for about a month before the EQ occurrence time. It is clear that the specific noise vanished shortly after the EQ occurrence.

The potential gradient data can be integrated along time in order to obtain the original piezoelectric field form generated in the focal area as a function of time (Thanassoulas 2007). The result of this transformation is shown in the following figure (8). At left (a) the resulted potential is presented while the red vertical bar indicates the EQ occurrence time. In the middle (b) the characteristic strain – stress function is presented while at right (c) the corresponding generated potential as a function of time due to (b) is presented. Following the basic laws of rock mechanics the rock formation will collapse closely to the upper non-linear part of (b) which corresponds to the upper part of (c). The figure (8, a) shows exactly that mechanism. The rock formation failure (EQ) took place at the upper part of figure (a) when it had reached very closely its fracture strain level suggested indirectly by the form, obtained by integration of the gradient data, of the piezoelectric potential.

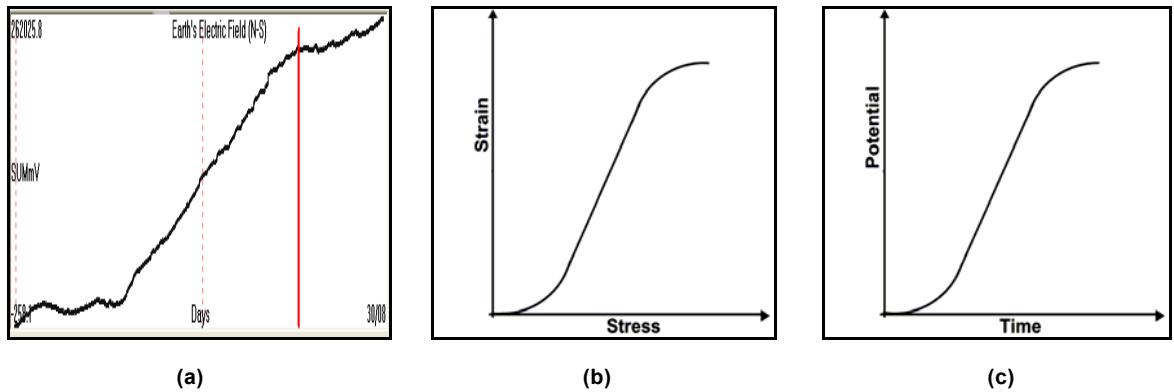

**Fig. 8.** Potential (a) generated by the EQ focal area, compared to the theoretical piezoelectric model (c) and the triggering stress inducing mechanism (b). Time duration of (a) is 35 days.

Following the presented model of figure (5, left) the gradient data were low-pass filtered in an attempt to identify the **VLP** piezoelectric signals which correspond to its first derivative in time. The results of that operation are presented in the following figure (9). A characteristic **VLP** (first derivative of the piezoelectric potential) signal that lasts for 20 days is clearly present that deviates from the background reference level.

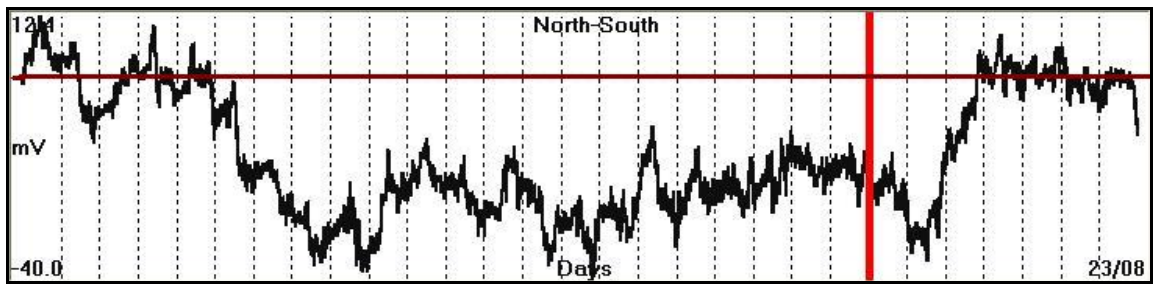

**Fig. 9. VLP** / first derivative of the piezoelectric potential identified after low-pass filtering of the raw data. The graph spans from 26/7/ to 23/8/1999 (20days). The brown horizontal line represents the background reference level while the red vertical bar indicates the occurrence time of the EQ.

It is characteristic that the EQ did occur almost at the end of that **VLP** signal conforming to the mechanism presented in figure (5, left) that suggests the occurrence of the rock formation fracture in the very near future.

If instead of using a low-pass filter we use a band-pass filter and select the pass band to be equal to 1 day, then it is possible to identify, due to the activation of the piezoelectric mechanism, the oscillating components of the earth's electric field which has been triggered by the tidally oscillating stress component (**K1, P1**) applied on the already highly strained seismogenic region (Thanassoulas et al. 1986, 1993). The latter is presented in the following figure (10). The figure (10) spans for the same period of time as the figure (9). It is very interesting to observe that the oscillating signal duration of figure (10) is almost the same as the **VLP** signal of figure (9). It is quite close to what is theoretically expected from figure (5, right). The piezoelectric mechanism justifies the simultaneous presence of the first derivative **VLP** signal and the oscillating one, since these are generated from the same part of the piezoelectric potential curve.

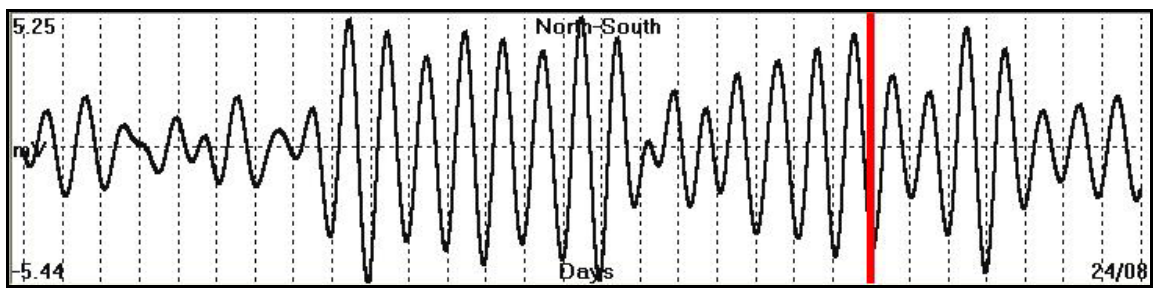

**Fig. 10.** Oscillating piezoelectric signal that spans from 26/7 to 23/8/1999. The signal lasts for 15 days and coincides quite well with the **VLP** of figure (9). The red vertical bar indicates the occurrence time of the EQ.

A smaller part of figure (10) of a few days is enlarged in order to compare the exact EQ occurrence time to the one of the amplitude peak of the oscillating electric field (Thanassoulas et al. 2011a). The enlarged graph is presented in figure (11).

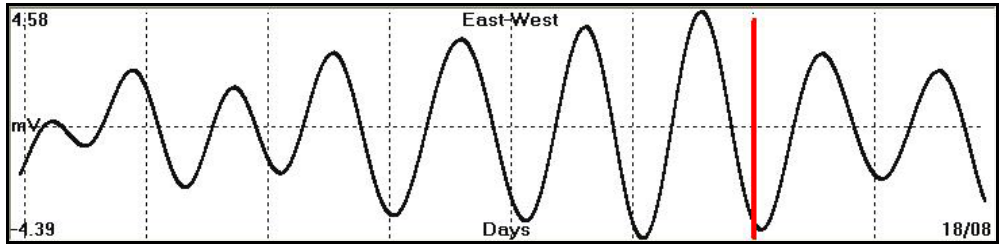

**Fig. 11.** Enlarged part of figure (10). The recording spans from August 11th to August 18th 1999. The EQ occurred 1hour and 22minutes before the closest electric field oscillation peak (Thanassoulas et al. 2011a).

So far, it has been demonstrated that preseismic electric signals of large periods (some hours – days) were generated before the Izmit EQ occurrence time by a hypothesized piezoelectric mechanism that was triggered at the seismogenic area. Furthermore, the very same mechanism justifies the generation of electric signals of frequencies ranging from KHz to MHz (Eftaxias et al. 2009). These high frequency electric signals can be generated at areas (**A**) and (**B**) of figure (6) left where rock fracturing is initiated. Therefore, due to the fact that an earthquake is the result of a gradual collapse of a rock formation, that evolves from small scale to large scale faulting, it is evident that during small scale fracturing high frequency **EM** waves can be generated. For the present case of the Izmit EQ that type of electric signals cannot be observed in the raw data recordings because the frequency response of the monitoring system is designed so that it rejects signals in the range of **KHz, ULF, MHz**. However, signals of larger period than a couple of minutes that are generated by the mechanism presented in figure (6, right), are clearly recorded. Some samples of that type of signals that preceded the Izmit EQ are presented in the following figures (12, 13, and 14).

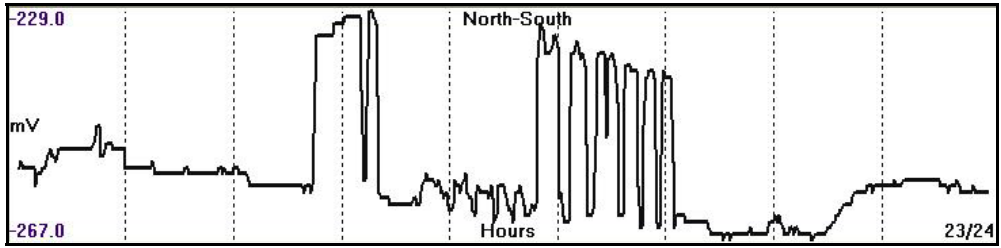

**Fig. 12.** Preseismic electric signal recorded at **VOL** monitoring site on 10/7/1999

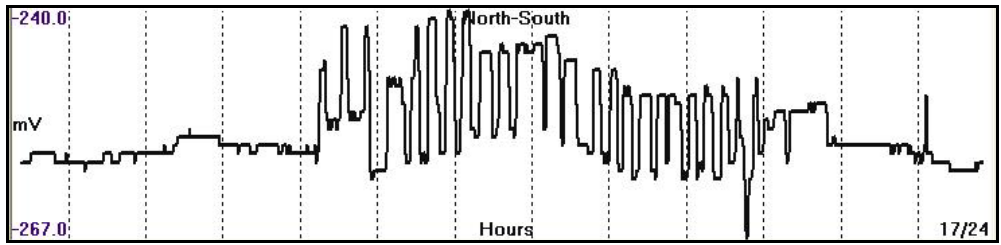

**Fig. 13.** Preseismic electric signal recorded at **VOL** monitoring site on 27/7/1999

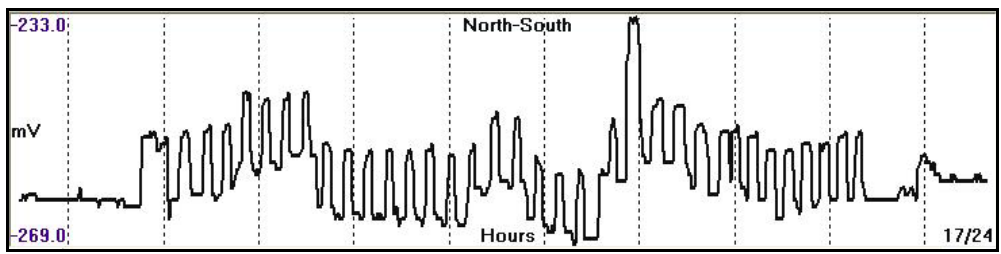

**Fig. 14.** Preseismic electric signal recorded at **VOL** monitoring site on 14/8/1999

The type of signals presented in figures (12, 13, and 14) was observed right from the start (20/6/1999) of the operation of the **VOL** monitoring site in digital mode. Therefore, we must accept that such signals were generated at the seismogenic region a few months before the Izmit EQ occurrence time.

Once the seismogenic area released its strain load through the Izmit EQ, it is expected that the piezoelectric signals should disappear. This is demonstrated in the following figures (15, 16).

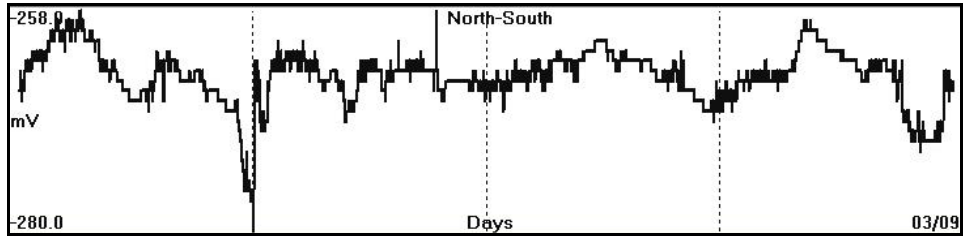

**Fig. 15. Earth's electric potential recorded at VOL monitoring site from 31/8 to 3/9/1999**

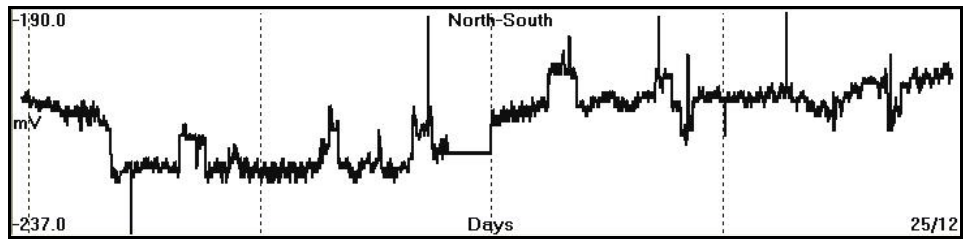

**Fig. 16. Earth's electric potential recorded at VOL monitoring site from 22/12 to 25/12/1999**

The absence of preseismic electric signals in both figures (15, 16) is very clear comparing to the figures (12, 13, and 14).

## 4. Discussion – Conclusions.

So far, all the theoretically expected electric signals that should be generated by a hypothetically triggered large scale piezoelectric mechanism in the focal area have been identified in or obtained from after processing the raw data of the VOL monitoring site recordings of the Earth's electric field. Therefore, it is justified to accept that the main physical mechanism that was triggered in the Izmit seismogenic region is the piezoelectric one.

During a six months period of recording of the Earth's electric field, only the NS component was recorded by VOL monitoring site. The following figure (17) corresponds to the recording day on 22/07/99. This recording was performed almost one month before the large Izmit, (17/8/1999, M=7.6) earthquake in Turkey and is more or less characteristic for the pre-earthquake recording period.

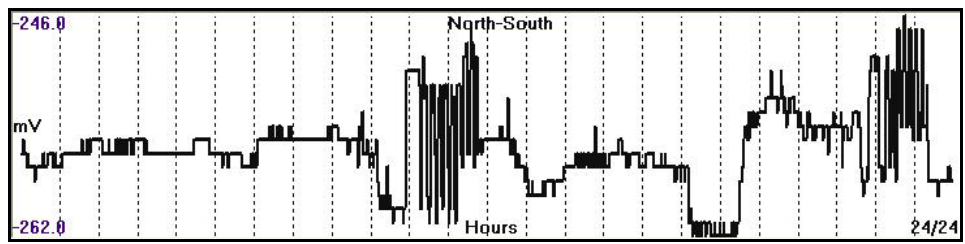

**Fig. 17. Preseismic electric signal recorded at VOL monitoring site on 22/7/1999**

An interesting feature of these signals is that, the start time of these signals coincides with two specific daytimes. The first one is around 9 a.m, while the second one is around 21.5 p.m. This observation was studied, in detail, by separating the "9.0 a.m" signals and the "21.5 p.m" signals in two groups (signal A, signal B). The following figure (18) represents the existence of electrical signals (signal A) as a function of time (in days), while the vertical axis represents the start-up time of each signal (in minutes) in the span of the day of its occurrence. In the horizontal axis of time, the earthquakes in Izmit, Athens and Duzce are marked with a red arrow.

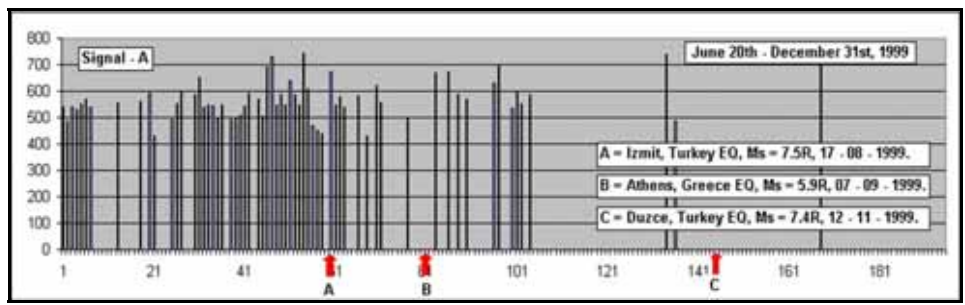

**Fig. 18. Daily presence of signals (A) is shown, for the period of time 20/06 – 31/12/1999.**

In the following figure (19), the signals (**B**) are presented with the same annotation.

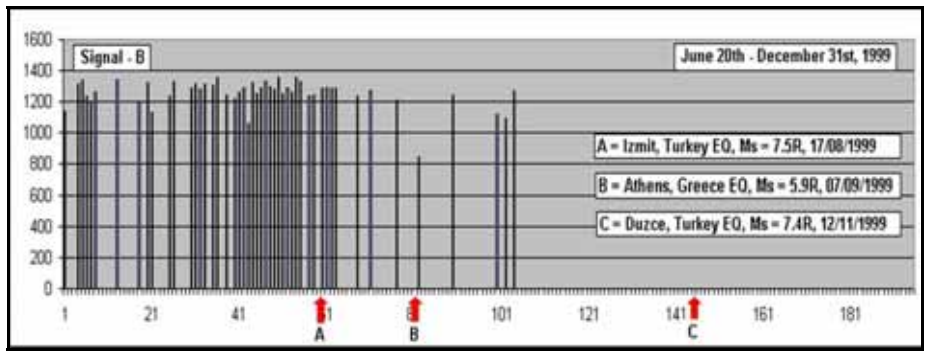

**Fig. 19. Daily presence of signals (B) is shown, for the period of time 20/06 – 31/12/1999.**

What is clear, from both figures, is the drastic decrease of the presence of the signals after the occurrence of Izmit EQ. On the other hand, before Duzce EQ, of a similar magnitude to Izmit EQ, no such signals were observed. This suggests that Izmit – Duzce regions may be considered as a unit seismogenic area, stress loaded and seismically activated. Consequently, the generated, electrical signals were produced by the entire, seismically active area and not only by the Izmit focal region. This is corroborated from the fact that Izmit – Duzce distance is of the order of 80Km (see fig. 20) which coincides quite well with the expected fracture length of the seismogenic fault, which is what is more or less expected for an EQ of **M = 7.6**.

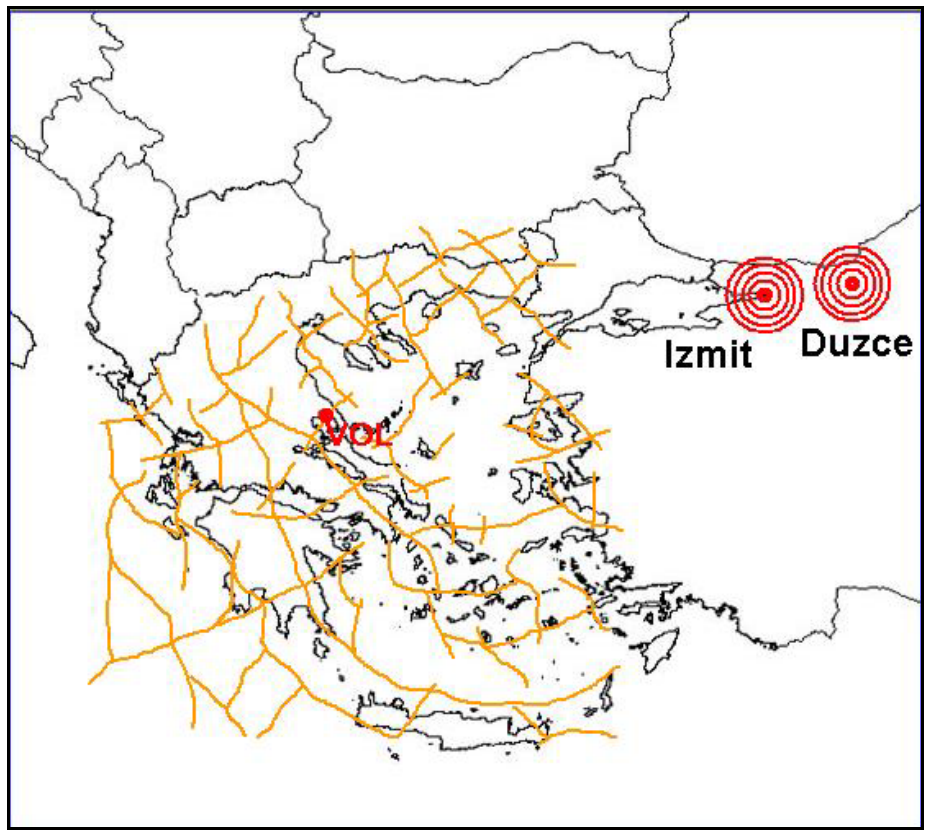

**Fig. 20. Izmit - Duzce location (red concentric circles) in relation to the VOL monitoring site in Greece.**

Therefore, when most of the stress-load of the entire seismogenic area had been released (by Izmit EQ), the rest of it was not capable of generating similar electrical signals. Viewing this pair of strong EQs from the point of view of electrical signals generation mechanism, it is characterized as a very interesting and spectacular, seismic event.

For both signals (**A, B**) the mean starting time has been calculated. For signals (**A**) the mean value (**MV$_A$**) was calculated as **569 minutes**.

$$MV_A = 569 \text{ minutes}$$

This corresponds to a mean starting time of 9hr 29 minutes. For signals (**B**), the same calculation results in a (**MV$_B$**) of **1257 minutes**. This corresponds to a mean starting time of 20hr 57 minutes.

$$MV_B = 1257 \text{ minutes}$$

**Finally, the mean time difference in time of occurrence of the electrical signals has been calculated, as:**

$$MV_B - MV_A = 11hr\ 28\ minutes.$$

Comparing this result to the Earth-tide components, its very close resemblance is revealed to (**K2**) (lunisolar) and (**S2**) (principal solar) components. A discrepancy of 4.17% has been calculated for the (**K2**) component, while a value of 4.4% corresponds to the (**S2**) one. The satisfactory results fit what was expected from the earlier theoretical analysis.

In simple words, the entire Izmit - Duzce seismogenic area was loaded at a such critical point of stress-strain charged conditions, so that it generated SES twice in a day, at both the amplitude peaks of the 24 hours lithospheric oscillation.

Since the lithosphere was set to a tidally triggered critical oscillating mode, it is justified to adopt the idea that other tidal components contributed to the triggering of the Izmit EQ. Therefore, the **M1** (Moon declination) tidal component that is calculated for the Izmit epicentre area is compared to the time of the EQ occurrence and presented in the following figure (21).

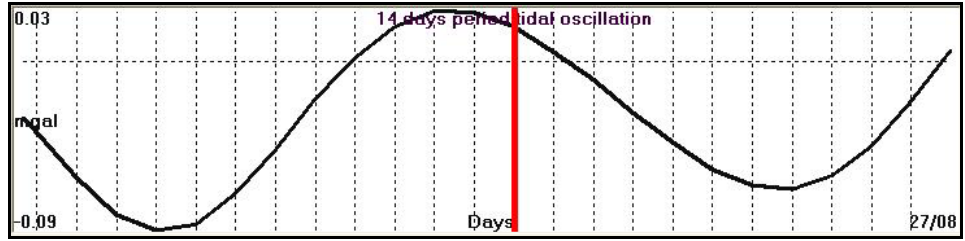

Fig. 21. Comparison of the Izmit EQ (17/8/1999) occurrence time (red bar) to the **M1** (Moon declination, black line) tidal component. The observed deviation **dt** of the occurrence time to the corresponding tidal amplitude peak is: **dt = 1.5 days**.

The Izmit EQ did not release the stress load that had been accumulated in the regional seismogenic area. Therefore, on 12/11/1999, another large EQ of Mw = 7.2 occurred, at 80 km eastern from Izmit, at Duzce area. That EQ is compared to the **M1** tidal component too. The following figure (22) shows this comparison.

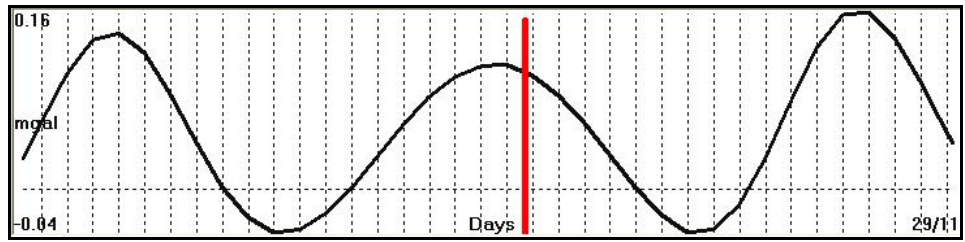

Fig. 22. Comparison of the Duzce EQ (12/11/1999) occurrence time (red bar) to the **M1** (Moon declination, black line) tidal component. The observed deviation **dt** of the occurrence time to the corresponding tidal amplitude peak is: **dt = 1.2 days**.

It is made clear from figures (21, 22) that the **M1** tidal component is the most probable mechanism that triggered the final fracture (EQ generation) of the Izmit - Duzce seismogenic region. The latter triggering mechanism was validated, during the preparation of this work, by another large EQ of Mw = 7.3 that did occur at South – Eastern Turkey on 23/10/11. A similar comparison of the calculated **M1** tidal component to the time of the EQ occurrence is presented in the following figure (23).

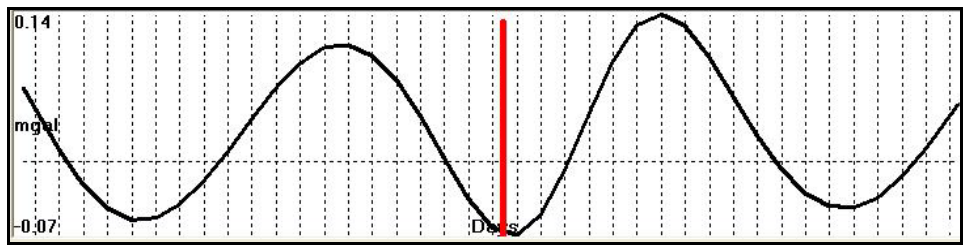

Fig. 23. Comparison of the South – Eastern Turkey EQ (Mw = 7.3, 23/10/2011) occurrence time (red bar) to the **M1** (Moon declination, black line) tidal component. The observed deviation **dt** of the occurrence time to the corresponding tidal amplitude peak is: **dt = .57 days**.

The latter results comply quite well to the conclusions of a global study of large EQs in relation to their local **M1** tidal component presented very recently by Thanassoulas et al. (2011).

A question that is obvious, after all that analysis, could be related to the predictability of the Izmit EQ. That is in simple words the possibility to know in advance, the location, time of occurrence and the magnitude of the referred EQ. These predictive parameters will be discussed separately each one as follows:

**Location:** In terms of regional seismological conditions what is known is the fact that the North Anatolian Fault (NAF) exhibits large seismic events along it, which large seismicity generally presents a west-ward drift. A mean 10-year period between triggering and subsequent rupturing shocks in the Anatolia sequence has been observed. Furthermore the study of the stress changes results in identifying areas (of known faults) of large probability for the occurrence of a large EQ in the next decade. Actually the Izmit regional seismogenic area was characterized by a large probability of 12% for the studied NAF segment (Sapanca fault, Stein et al. 1997). Therefore, the Izmit EQ, of 1999, validated the Stein et al. (1997) analysis. In a recent seismological analysis of the same seismic event it is revealed that the earthquake was preceded by a seismic signal of long duration that originated from the hypocenter. The signal consisted of a succession of repetitive seismic bursts, accelerating with time, and increased low-frequency seismic noise. These observations show that the earthquake was preceded, for 44 minutes, by a phase of slow slip occurring at the base of the brittle crust. This slip accelerated slowly initially, and then rapidly accelerated in the 2 minutes preceding the earthquake (Bouchon et al. 2011). The latter seismic signal could be considered as a probable short-term seismic precursor since the Izmit regional area was highly stress loaded and consequently that signal was predicting the area to be ruptured in the very near future.

On the other hand, the recorded in Greece preseismic electric signals cannot provide the future epicentre area for the same seismic event due to the fact that triangulation of the Earth's registered electric field requires at least two distant monitoring sites (Thanassoulas, 2007).

**Time of occurrence:** As far as it concerns the time of occurrence of the Izmit EQ, following the Stein et al. (1997) study an estimate can be made for a future EQ within the next few years to come after the last NAF rupture that preceded the Izmit EQ. That kind of prediction refers to a medium term one. On the other hand, Bouchon et al. (2011) seismic signal was "a posteriori" identified and within a very short time period (44 minutes) before the Izmit EQ occurrence. Therefore, it was "too short-term" for practical warning purposes in order to avoid people deaths. In order to estimate the time of occurrence of the Izmit EQ the registration of the Earth's electric field combined to the M1 tidal lithospheric oscillation can provide valuable information concerning the future EQ time of occurrence. The postulated methodology is presented in the following figure (23). What is really required is: to identify the last phase of the lithospheric deformation when the rock formation failure will take place and, moreover, during this last phase of deformation, the time of occurrence of the maximum fine stress oscillation (induced by M1) which will actually trigger the EQ. The black line represents the generated piezoelectric potential as a function of time. At its upper part (right) the potential curve bends thus showing that the seismogenic area is very close to rupturing time. The green line represents the M1 tidal component. The red bar indicates the time of occurrence of the Izmit EQ. The coincidence of the time of occurrence of the EQ, with the upper bend of the piezoelectric potential and with the maximum of the lithospheric stress load oscillation is pretty obvious.

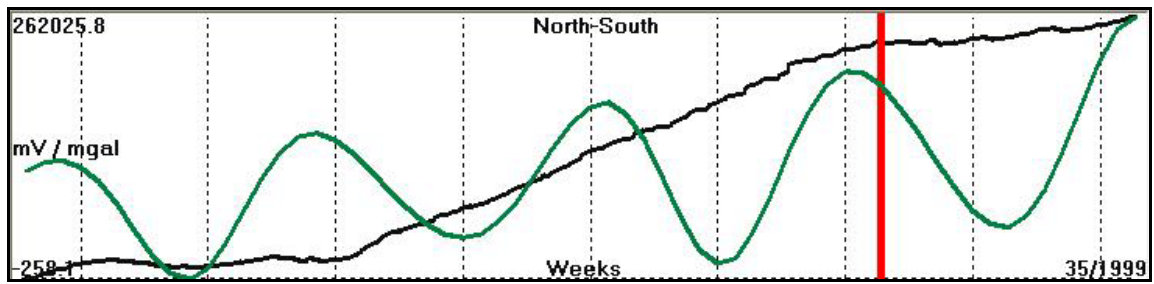

**Fig. 23.** Correlation of the piezoelectric potential (black line) observed in Greece (VOL monitoring site) before the Izmit EQ with the maximum amplitude (thus maximum stress-load) of the lithospheric oscillation (green line) observed at the epicentral area due to the M1 tidal component, and with the time of occurrence (red bar) of the Izmit EQ.

Very similar results to the ones shown in figure (23) have been presented by Thanassoulas et al. (2008a, 2009, 2010) after analysing the earth's electric field that was registered before the occurrence of some large EQs, in Greece.

**Magnitude:** During the occurrence of an earthquake the dynamic strain energy that is stored in the seismogenic region is released as a mechanical kinetic one. The amount of the released energy is expressed by a number that defines the magnitude of the seismic event. Consequently, it is possible to know in advance the maximum magnitude of an earthquake that is possible to occur at any seismogenic area provided that the stored strain energy has been calculated. Therefore it is required that either the "Lithospheric Seismic Energy Flow Model LSEFM" (Thanassoulas, 2007) will be applied on an already known specific seismogenic area which is just before failure, or the corresponding regional seismic potential maps are available, compiled from the past (some years) seismicity and indicate the expected maximum EQ magnitude. For the present case of the Izmit EQ none of the latter could be applied before its occurrence, due to the lack, by that time (1999), of the appropriate methodology and the corresponding seismic potential maps.

In conclusion, the Izmit EQ, due to its large magnitude and due to the large amplitude generated preseismic electric signals, could have quite accurately been predicted, provided that a widely spread (along NAF) network of monitoring stations of the earth's electric field had been deployed well in advance, for continuously monitoring of the Earth's electric field.

## 5. References.